\documentclass[onecolumn]{article}

\usepackage{graphicx}
\usepackage{amsmath}
\usepackage{amssymb}
\usepackage{bigstrut}
\usepackage{tabularx,ragged2e,booktabs,caption,authblk}
\newcolumntype{C}[1]{>{\Centering}m{#1}}

\everymath{\displaystyle}
\usepackage{fullpage}
\usepackage{cite}
\usepackage{float}
\usepackage{url}

\begin{document}

\title{Modeling human activity-related spread of the spotted lanternfly (\emph{Lycorma delicatula}) in the US}

\author{Daniel Str\"{o}mbom$^1$\thanks{\mbox{Corresponding author: stroembp@lafayette.edu}} , Autumn Sands$^1$, Jason M. Graham$^2$, Amanda Crocker$^1$, Cameron Cloud$^1$, Grace Tulevech$^1$, Kelly Ward$^1$.\\\small{$^{1}$Department of Biology, Lafayette College,  Easton, PA, USA.}\\\small{$^{2}$Department of Mathematics, University of Scranton, Scranton, PA, USA.}}

\date{}
\bigskip
\maketitle

\begin{abstract}
The spotted lanternfly (\textit{Lycorma delicatula}) has recently spread from its native range to several other countries and forecasts predict that it may become a global invasive pest. In particular, since its confirmed presence in the United States in 2014 it has established itself as a major invasive pest in the Mid-Atlantic region where it is damaging both naturally occurring and commercially important farmed plants. Quarantine zones have been introduced to contain the infestation, but the spread to new areas continues. At present the pathways and drivers of spread are not well-understood. In particular, several human activity related factors have been proposed to contribute to the spread; however, which features of the current spread can be attributed to these factors remains unclear. Here we collect county level data on infestation status and four human activity related factors and use statistical methods to determine whether there is evidence for an association between the factors and infestation. Then we construct a mechanistic network model based on the factors found to be associated with infestation and use it to simulate local spread. We find that the model reproduces key features of the spread 2014 to 2021. In particular, the growth of the main infestation region and the opening of spread corridors in the westward and southwestern directions is consistent with data and the model accurately forecasts the correct infestation status at the county level in 2021 with $81\%$ accuracy. We then use the model to forecast the spread up to 2025 in a larger region. Given that this model is based on a few human activity related factors that can be targeted, it may prove useful in informing management and further modeling efforts related to the current spotted lanternfly infestation in the US and potentially for current and future invasions elsewhere globally.
\end{abstract}

\section*{Introduction}
The spotted lanternfly (\textit{Lycorma delicatula}) is an insect native to China, India and Vietnam that is currently considered an invasive pest in South Korea, Japan and the United States \cite{Dara2015,Jung2017}. In regions where it is invasive, it is a threat to many naturally occurring plants and a number of economically important interests including the grape, orchard and lumber industries \cite{Barringer2015,Urban2023}. 

The spotted lanternfly was first confirmed in the US in 2014 and has since rapidly spread and expanded its distribution \cite{NYSIPM2020}. Quarantine zones and permitting by respective state Departments of Agriculture have been implemented to restrict the movement of certain products and any living life stage of the spotted lanternfly in an attempt to limit the spread to new regions \cite{Urban2020}. In areas where the lanternfly is already present a number of control measures, including insecticides \cite{Leach2019}, parasite-based biological controls \cite{Liu2019a,Lee2019}, traps \cite{Francese2020}, egg scraping \cite{Cooperband2018}, and removal of the primary host plant \cite{Aphis2018} have been applied. However, these efforts appear to be insufficient given that the lanternfly is still spreading rapidly to new regions \cite{Cook2021,Urban2023}. In addition, earlier published niche models based on climate and other factors forecast that the spotted lanternfly may eventually establish itself in a much larger portion of the US than it currently occupies and elsewhere globally \cite{Jung2017,Wakie2020}. However, these models only forecast the eventual maximal distribution, they do not provide information about how the lanternfly is likely to spread from its current limited distribution to this maximal distribution \cite{Hao2019} nor what the key drivers of the local spread are. To address the issue of forecasting how the spotted lanternfly, and other pests and pathogens, might spread over time a species agnostic dynamic spatio-temporal forecasting system was recently introduced \cite{Jones2021}. Unlike previous models for the lanternfly infestation in the US this system can also provide estimates for when the lanternfly is expected to arrive in a particular region, rather than just that it eventually might. For example, it has been used to forecast that there is a high chance that California will have lanternfly infestations in 2033 \cite{Jones2022}. While this system represents a significant advance in dynamic forecasting its high complexity and data-driven nature makes it less suitable for obtaining a causal understanding of the current local spread in terms of key drivers, and how to limit the spread by targeting these. To address such questions a mechanistic model based on a limited number of factors may be more appropriate \cite{Baker2018}.

The spotted lanternfly is known to spread through two distinct processes; short distance dispersal and long distance dispersal \cite{Lee2019}. Short distance dispersal includes natural flight capabilities of up to 40m to infest new host plants \cite{Baker2019} and/or local human assisted transport \cite{Wolfin2019}. Long distance dispersal refers to establishment of satellite populations away from the main area of infestation \cite{Lee2019}, and several potential pathways and drivers of long distance dispersal in the spotted lanternfly have been proposed in the literature. In particular, a number of human activity related factors including the transportation of packing materials and vehicles \cite{Dara2015,Barringer2015}, transportation of materials related to gardening and plants \cite{Hong2012}, and human activity/population density \cite{Lee2019,Cook2021,Namgung2020,Ladin2023}. Unlike climate, geography, host plant distribution, temperature, and other non-human activity related factors that are known, or believed, to affect the spread and settlement of the lanternfly, these human activity related factors may be more readily targetable and affectable by management efforts. A minimal model based on such factors capable of replicating key features of the actual spread could be used as a framework for assessing proposed management efforts before they are deployed, and potentially even to help determine what they should be.

\section*{Results}
\begin{figure}[!h]
\centering
\includegraphics[width=0.9\columnwidth]{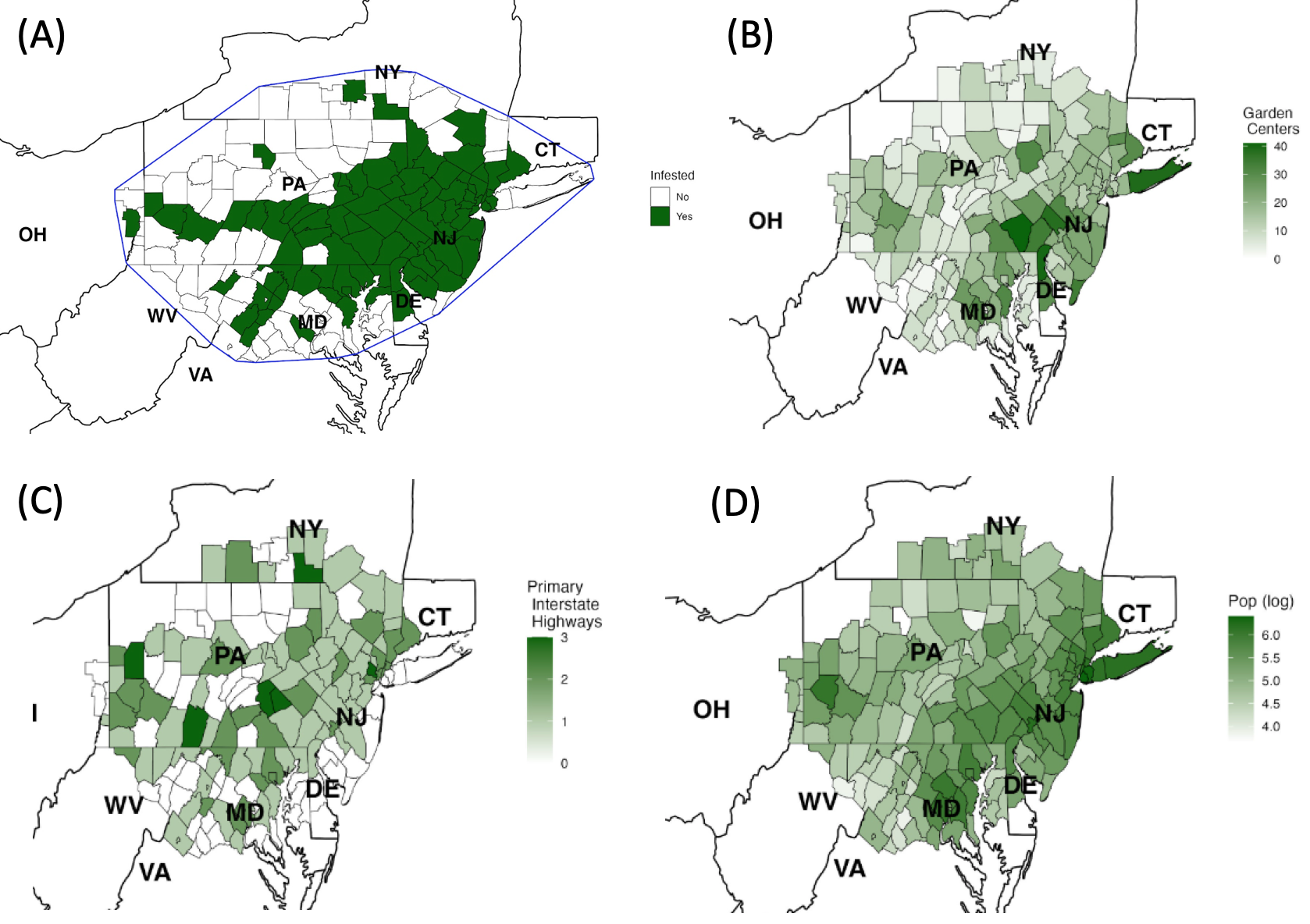} 
\caption{{\bf Visualization of the assessment region and the data collected.} (A) County map showing the counties designated as infested by the New York State Integrated Pest Management \cite{NYSIPM2020} on September 28th 2021 and the 166 county region defined as the convex hull of those infested counties. (B) The number of garden centers present in each county. (C) The number of primary interstate highways transecting each county. (D) The population of each county on a log scale.}
\label{fig:1}
\end{figure}

The data collected for a 166 county region that contained all infested counties \cite{NYSIPM2020} on September 28th 2021 (Fig \ref{fig:1}A), including number of garden centers (Fig \ref{fig:1}B), primary interstate highways (Fig \ref{fig:1}C) and population (Fig \ref{fig:1}D) is presented in Figure \ref{fig:1}.

By fitting a generalized additive logistic regression model \cite{rcore,wood2003,wood2017generalized} we find that the data provides evidence for the increase in likelihood of infestation in a county with the presence of interstate highways and larger populations (Table \ref{tab:1}). Additionally, the analysis also suggests that the data provides evidence for the increase in likelihood of infestation in a county with an increase in the number of garden centers only when there is also the presence of a primary interstate highway in that county.\\
See the Methods section for information about the data collection and the statistical methods used.

\begin{table}[h!]
\caption{\label{tab:1}Summary table for estimates for linear terms in generalized additive
model. Results are on the log-odds scale. }\tabularnewline

\centering
\begin{tabular}{l|r|r|r}
\hline
Coefficient & Estimate & SE & p-value\\
\hline
IS Presence - Yes & 1.21 & 0.63 & 0.054\\
\hline
Garden Centers & -1.29 & 0.63 & 0.041\\
\hline
Population (log) & 1.68 & 0.84 & 0.046\\
\hline
IS Presence - Yes : Garden Centers & 1.96 & 0.69 & 0.005\\
\hline
\end{tabular} \label{tab:1}
\end{table}

\subsection*{Model}
Based on insights presented in the literature and the results of our statistical analyses above we construct a network model of the spread consisting of two components. General short distance spread on the network of adjacent counties and long distance spread on the primary interstate highway network with garden center and population density dependent infestation probabilities. The components and parameters of the model are described in Table \ref{tab:2}.

\begin{table}[h!]
\caption{Components and parameters of the model.}\tabularnewline
\centering
\begin{tabular}{cl}
Symbol       & Description 
\\
\hline
\\
$t$       & Timestep (1 year).                                                                                                                                                                                                                                   \\
$i,j$     & County indices.                                                                                                                                                                                                                                      \\
$\bar{q}_t$     & Infestation vector. $q_{t,i}=1$ if county $i$ is infested at time $t$, and $q_{t,i}=0$ if it is not.                                                                                                                                                 \\
$A$       & \begin{tabular}[c]{@{}l@{}}Adjacency matrix encoding the network of adjacent counties. \\ $a_{ij}=1$ if county $i$ and county $j$ share a border, and $a_{ij}=0$ if they do not.\end{tabular}                                                    \\
$H$       & \begin{tabular}[c]{@{}l@{}}Adjacency matrix encoding the network of counties connected by a primary\\ interstate highway. $h_{ij}=1$ if county $i$ and county $j$ share a highway, and\\ $h_{ij}=0$ if they do not.\end{tabular} \\
$p$       & \begin{tabular}[c]{@{}l@{}} Probability of spread to an uninfested county from each of its adjacent infested\\ counties. Estimated using the 2014-20 data. See the Methods section for details.\end{tabular}                                                               \\
$\bar{s}_x$     & \begin{tabular}[c]{@{}l@{}}Establishment probability vector for primary interstate highway spread by the\\ factor x. Ex. $\bar{s}_{x}$=(number x in county $i$)/(total number x). We consider two such \\factors here: garden centers ($\bar{s}_g$) and population ($\bar{s}_p$).
\end{tabular}              
\end{tabular}

\label{tab:2}
\end{table}

The spread on each of the two interconnected networks follow the independence model introduced in Leung et al. 2004 (See equation (1) in \cite{Leung2004}) modified to include the garden center and population dependent spread on the primary interstate highway network. More specifically, the spread on the adjacency network is given by
\begin{equation}\label{eq:1} \Pr\{q_{i,t+1}=1 | q_{i,t}=0 \} =1-(1-p)^{\sum_jA_{ij} q_{j,t}},\end{equation} 
and the garden center dependent spread on the primary interstate highway network by
\begin{equation}\label{eq:2} \Pr\{q_{i,t+1}=1 | q_{i,t}=0 \} =1-(1-s_{g,i})^{\sum_jH_{ij}q_{j,t}}, \end{equation}
and the population dependent spread on the primary interstate highway network by
\begin{equation}\label{eq:3} \Pr\{q_{i,t+1}=1 | q_{i,t}=0 \} =1-(1-s_{p,i})^{\sum_jH_{ij}q_{j,t}}. \end{equation}

We implemented the combined modes of spread over the interconnected adjacency and primary interstate highway networks in Matlab and analyzed the model via simulations. See the Methods section for details on model construction, parameterization, implementation, simulation and assessment.

\begin{figure}[ht]
\centering
\includegraphics[width=\columnwidth]{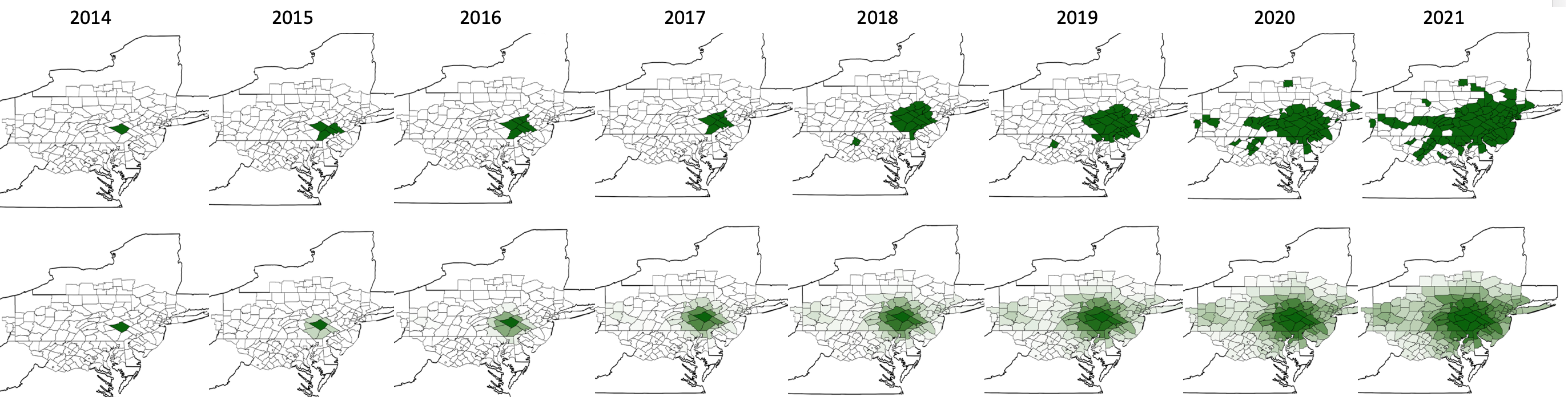} 
\caption{{\bf Comparison of observed and model forecasted infestation in the 166 county model 2014-21.}
The Top row displays the actual observed infestation data from 2014-21. The Bottom row displays the corresponding model forecasts for 2015-21. The coloring indicates the proportion of simulations in which a county became infested in a specific year according to the model. More specifically, a county is solid green if all 1000 simulations rendered it infested in a particular year and white if none of the 1000 simulations resulted in it being infested.}
\label{fig:2}
\end{figure}

The results of simulating the model in the 166 county region starting with Berks county, PA, as the only infested county in 2014 and recording the proportion of simulations that yielded each county infested in each year 2015-2021 is presented in the bottom row of Figure \ref{fig:2}. We note that the model constructed based on the included factors reproduces key characteristics of the observed spread both qualitatively and quantitatively. Comparing the Top and Bottom rows in Figure \ref{fig:2} shows that not only is the model forecasted growth of the infested region relatively consistent with the observed over the years, but the model also forecasts the opening up of spread corridors from the main infestation region around Berks County, PA, west toward the Pittsburgh area and south-southwest through VA. In particular, the model starts suggesting already in 2017-18 that Alleghany County, PA, had an increasing risk of infestation before it was finally designated as infested in 2020. Similar observations are made in each main direction of infestation spread. Quantitatively, using the rule that if more than half of the simulations resulted in a county being infested to designate a county as infested according to the model, the model predicts the correct 2021 county infestation status with $81\%$ accuracy. More specifically, it forecasts the infestation correctly in 135 counties and incorrectly in 31 counties. Of the 31 mismatches there are 16 false positives, where the model suggests the county should be infested but the data disagrees, and 15 false negatives, where the model suggests the county should not be infested but the data says it is.

\begin{figure}[h!]
\centering
\includegraphics[width=\columnwidth]{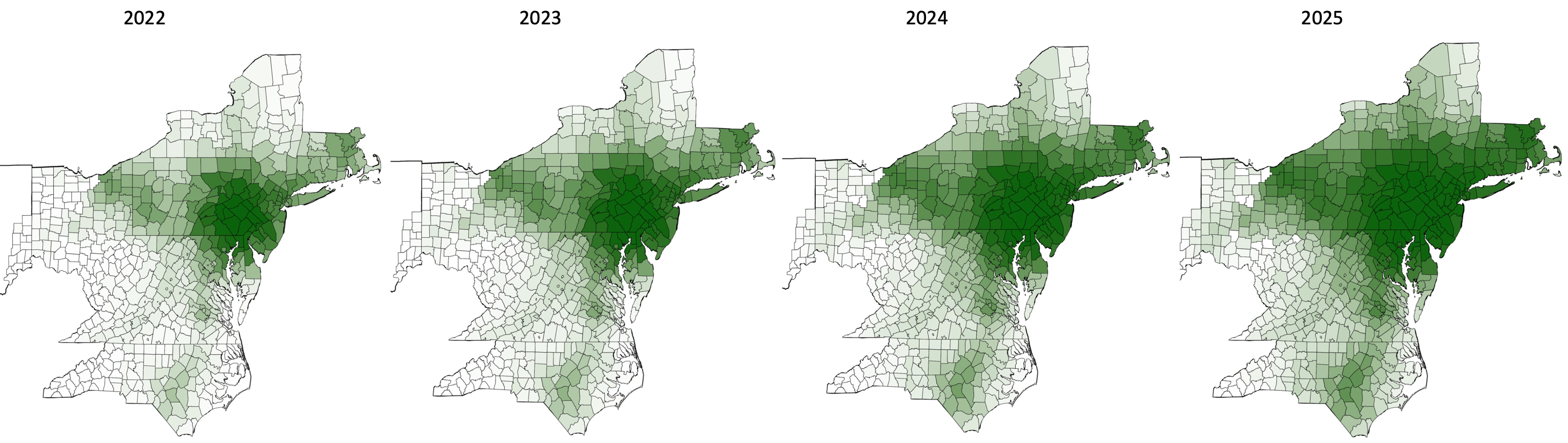} 
\caption{{\bf Model forecast for 2022-25 in the 581 county region}
We note that the model suggests that the corridors of spread west, southwest, and south will all intensify over the years. In particular, the model suggests that the area around Cleveland, OH, has had an increased and increasing risk of infestation at least since 2021, and it was recently labeled as infested by the New York State Integrated Pest Management Program. The model also forecasts that the infestation in NC will increase substantially over the next few years.}
\label{fig:3}
\end{figure}

Once the model had been assessed in the 166 county region the same parameterized model was simulated in a larger 581 county region. These simulations in the larger region forecasts that the spread along the western corridor through OH and IN, the southwestern corridor on the border of WV and VA, and the southern corridor through VA into NC, will intensify up to the year 2025 (Figure \ref{fig:3}).

See the Methods section for more detailed descriptions of the model, model assessment, and the simulation protocols. 

\section*{Discussion}
The computational model created based on the human activity related factors; primary interstate highways, garden centers and population reproduce both qualitative and quantitative features of the spread through 2014-21. Using the computational model over a larger region of the US to forecast the spread of infestation beyond 2021 suggests that spread will accelerate over this period.

Despite being based on only three specific statistically supported human activity related factors, our model not only reproduces qualitative features of the actual spread, but also forecasts the correct infestation status for 81\% of the counties included in the assessment region in 2021. This, together with the observed infestation pattern itself (Figure \ref{fig:2} Top row) supports the assertion that the primary interstate highway system may be a critical pathway for the long distance dispersal of the spotted lanternfly in the US and that human activity in general, and business dealing in plants and plant related materials in particular, are associated with the dispersal along this pathway. While this is consistent with findings and propositions by others as described previously, we note that our finding that primary interstate highways are associated with infestation, rather than road density in general \cite{Cook2021} or even all interstate highways, may be a particularly useful insight. There are relatively few primary interstate highways which may make targeting for management and modeling based on those easier and less complex than if a larger classes of roads and transportation pathways are considered. In addition, our statistical finding that garden centers only appear to be associated with increased risk of infestation when primary interstate state highways are present may be useful for prioritizing monitoring and/or management efforts involving garden centers.

Our model does not take into account climate, geography, host plant distribution, temperature, and other factors that are known, or believed, to affect the spread and settlement of the lanternfly. This may explain why the model forecasts the incorrect infestation status in $19\%$ of included counties. If a factor we do not include is severely disadvantageous to lanternfly infestation that factor could explain some of the 16 false positives, and if the excluded factor is severely advantageous to lanternfly infestation it could explain some of the 15 false negatives. However, other explanations for, in particular, the false positives do exist. Notably that sampling effort by authorities and reporting by the public is not uniform throughout the region under consideration \cite{Cook2021} so lanternfly infestations may be present in a county but not reported. While it may eventually be useful to add more factors and/or couple our model with niche models \cite{Jung2017,Wakie2020} and/or dynamic forecasting system \cite{Jones2021,Jones2022} at present it may be most useful in its current minimal form. Primarily because it is already reasonably effective and adding further elements would increase its complexity and reduce its ability to elucidate the causal effects of the specific human activity related factors on the spread of infestation. However, once the infestation has progressed further, both the niche models and the dynamic forecasting system suggest that the conditions for successful establishment of infestation drops significantly (Figure 5 in \cite{Jones2022}). If those forecasts are correct our model will produce false positives in these new regions because it does not take into account climate and other factors that they do. If/when the infestation reaches areas with low risk of infestation it will not only serve as a test of our model but also both the niche models \cite{Jung2017,Wakie2020} and forecasting system \cite{Jones2021,Jones2022}. If the spread significantly slows down or stops our model needs revision, but if it proceeds at the same rate in designated low risk areas or move into areas forecasted as unlikely to be suitable the niche models and forecasting system, and potentially our model too, need revision. 

Our forecasts in the larger region suggest that the lanternfly infestation will continue its long range dispersal along the primary interstate highway system in all directions and short distance dispersal will drive the spread outward from the new satellite infestations along these routes (Figure \ref{fig:3}). This information may be useful on the ground as a complement to the forecasting system in \cite{Jones2022} to forecast where lanternfly infestation is likely to occur next when planning for the application of control measures. In part, because preventing infestation or managing an infestation early is preferable to dealing with a full blown infestation later on \cite{Dent2020}, but also because of the findings in \cite{Strombom2021}. Based on available data on the life-stage survival and fecundity \cite{Strombom2021} estimated that the average annual growth rate of the US spotted lanternfly population is about 5.47. Given this rapid growth even a control measure that kills 100\% of treated lanternfly require at least 35\% of all lanternfly present in all stages to be treated to induce even the slightest decline in the annual population and this percentage increases rapidly for less effective controls. Given that resources are limited this suggests that using them in the right locations may be critically important. For example, our model forecasts that over the next few years the infestation will intensify in North Carolina (Figure \ref{fig:3}). This state has become one of the largest wine producing states with over 525 individually owned vineyards in 2017 \cite{Winslow2017} and in previously infested regions the grape industry has been particularly adversely affected with one vineyard in Pennsylvania reporting a 90\% loss in yield \cite{Urban2020}. Our work, including both model and statistical analysis, in conjunction with the infestation data suggests that spread occurs along the primary interstate highways, so efforts to limit the spread risk southward along primary interstates 77, 85, and 95 may lower the risk for North Carolina's wine industry. We are aware that the Pennsylvania Department of Agriculture are focusing on transportation corridors such as major highways and tourist destinations \cite{Powers2021}, but we are unable to find information about which specific highways or areas they are targeting and/or what decisions to target these are based on. Our model, or a mechanistic model similar to it, could help inform such decisions. Whether it is to assess two or more competing management alternatives with given estimated county-level impacts before implementation, or to minimize spread to a particular region, or even to minimize the overall spread. Comparing alternatives will be particularly straightforward since that would amount to changing county level parameters. For example, say that some action is estimated to decrease the long-distance dispersal through a county by 30\%, then the weights on the network links from that county would be reduced by 30\% and new simulations run to observe the outcome. Minimizing spread to a particular region, or minimizing the overall spread, requires more work, but there is a wealth of analysis tools available for entities that spread on networks (e.g. infections \cite{Nandi2016,Enns2012}) that could be used to rank the nodes (counties) and the links (transmission pathways) based on their contribution to the overall spread. We believe that using mechanistic modeling as a complement to other approaches to estimate where the limited resources available to combat the infestation should be deployed for maximum impact with respect to whichever objective is set by professionals on the ground will be necessary to significantly reduce the spread of lanternfly in the US. The low-complexity model based on statistically established factors we present here represents one step in this direction.


\section*{Methods}

\subsection*{Data collection} \label{sec:data}
First, we defined a region for our statistical analyses and model assessment such that it is plausible that the lanternfly would have spread to and infested the counties in this region if the conditions had been favorable by following the convex hull approach used in \cite{Cook2021}. This method led to a 166 county region as the convex hull of the counties designated as infested by the New York State Integrated Pest Management \cite{NYSIPM2020} on September 28th 2021 (Fig \ref{fig:1}A). We also collected data on the number of garden centers (Fig \ref{fig:1}B), primary interstate highways (Fig \ref{fig:1}C) and population (Fig \ref{fig:1}D) for each county in the region. More specifically, for each county in the 166 county region we collected
\begin{enumerate}
	\item{The names and FIPS codes of its adjacent counties from the United States Census Bureau \cite{USCB2020}.}
	\item{The year in which it was designated as infested by consolidating information from the New York State Integrated Pest Management \cite{NYSIPM2020}, the Pennsylvania Department of Agriculture \cite{PDA2015,PDA2016,PDA2017,PDA2019}, State of New Jersey Department of Agriculture \cite{Wolfe2018}, and the Virginia Polytechnic Institute and State University \cite{VPISU}.}
	\item{The number of businesses designated as garden centers (retail or wholesale). This data was collected in October 2021 by manually searching "Garden centers in [county name] county" in Google Maps \cite{GoogleMaps2021} and verifying that each search result was indeed a garden center that lies within the county boundary. (Figure \ref{fig:1}B).}
 \item{The 2019 population estimate from the United States Census Bureau \cite{USCB2019}. (Figure \ref{fig:1}D).}
  \item{The primary (two-digit) interstate highways that transect it and their identification numbers from Google Maps \cite{GoogleMaps2021} (Figure \ref{fig:1}E).}
\end{enumerate}

In the larger 581 county region we collected the same data using the same sources. The exception being the infestation year because no county outside the 166 county region was infested at that time.

\subsection*{Statistical analysis to determine relevance of factors}\label{sec:stats}

\begin{figure}
{\centering 
\includegraphics[width=0.65\columnwidth]{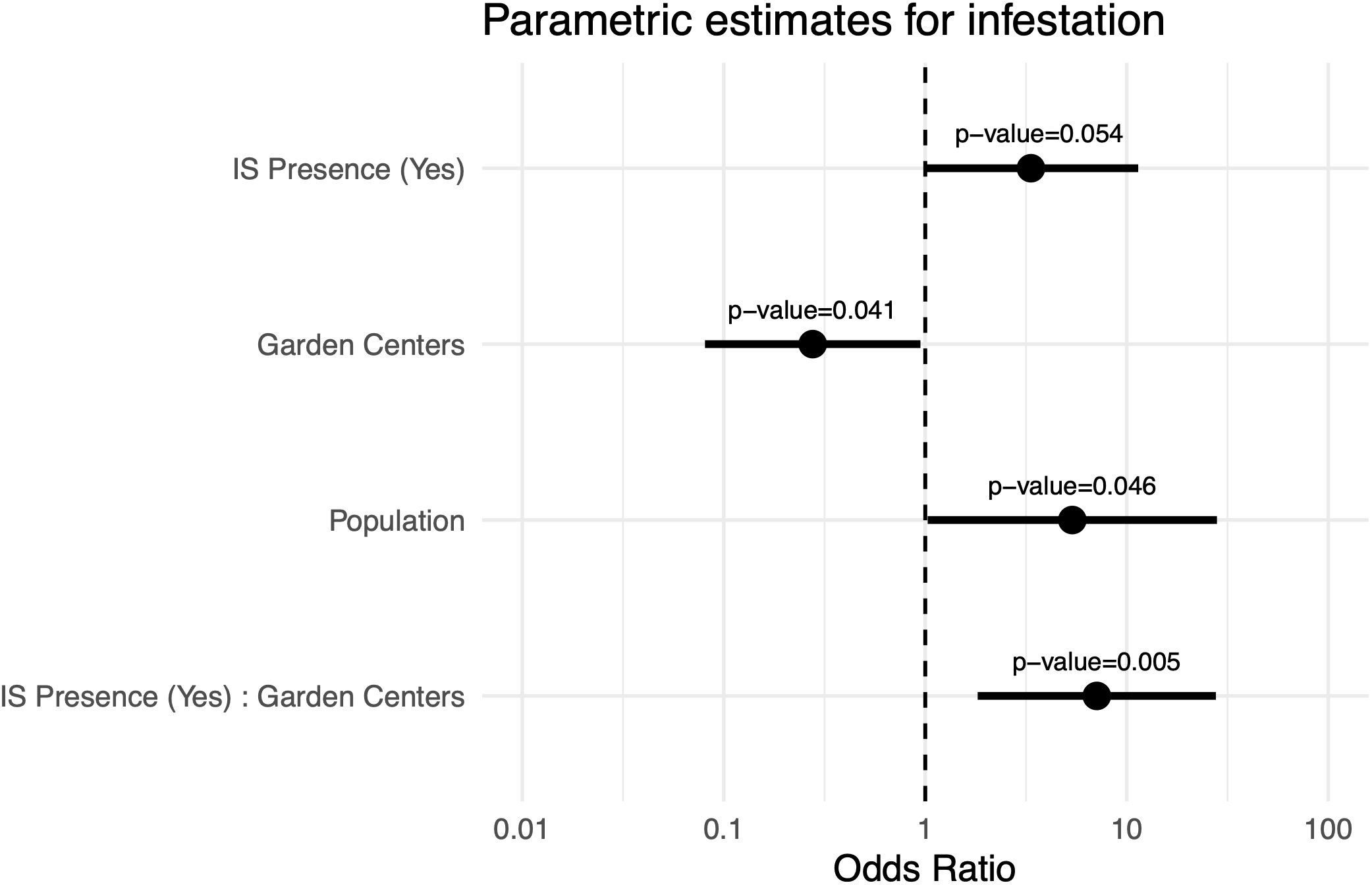}
\caption{\label{fig-estimates}\textbf{Estimated coefficients with 95\% CIs for the parametric terms in the generalized additive model}. Results are on the scale of odds-ratio. The likelihood of infestation for a county in 2021 is predicted to increase with the presence of interstate highways (IS) and larger populations. Interestingly, the likelihood of infestation for a county in 2021 is predicted to increase with an increase in the number of garden centers only when there is also the presence of a primary interstate highway for that county.}}
\end{figure}

For a statistical test of our hypotheses we used R version 4.3.1 and package \texttt{mgcv} version 1.9 to fit a generalized additive
logistic regression model \cite{rcore,wood2003,wood2017generalized}. Specifically, we model the log-odds of county infestation in 2021 with tensor product smoothing for longitude and latitude to control for spatial-autocorrelation and include parametric terms for presence/absence of two-digit interstate highway, number of garden centers, county population, and an interaction term for presence/absence of primary (two-digit) interstate highway and number of garden centers. To deal with issues of convergence and variables of different scale, we log transformed the population and normalized the number of garden centers. We used the packages \texttt{DHARMa} version 0.4.6 and \texttt{gratia} gratia version 0.8.1.34 for diagnostics to assess model assumptions \cite{dharma2022,gratia2023}. Figure~\ref{fig-estimates} shows the estimates on the odds-ratio scale for the parametric terms in our generalized additive logistic regression, as a complement to Table~\ref{tab:1} which displays the values for the same estimated coefficients on the log-odds scale.

\subsection*{Model construction, parameterization, implementation, simulation and assessment}
The model was implemented in Matlab and here we describe the main model related elements in more detail.

\subsubsection*{Model spaces}
\begin{figure}[ht]
\centering \includegraphics[width=0.8\columnwidth]{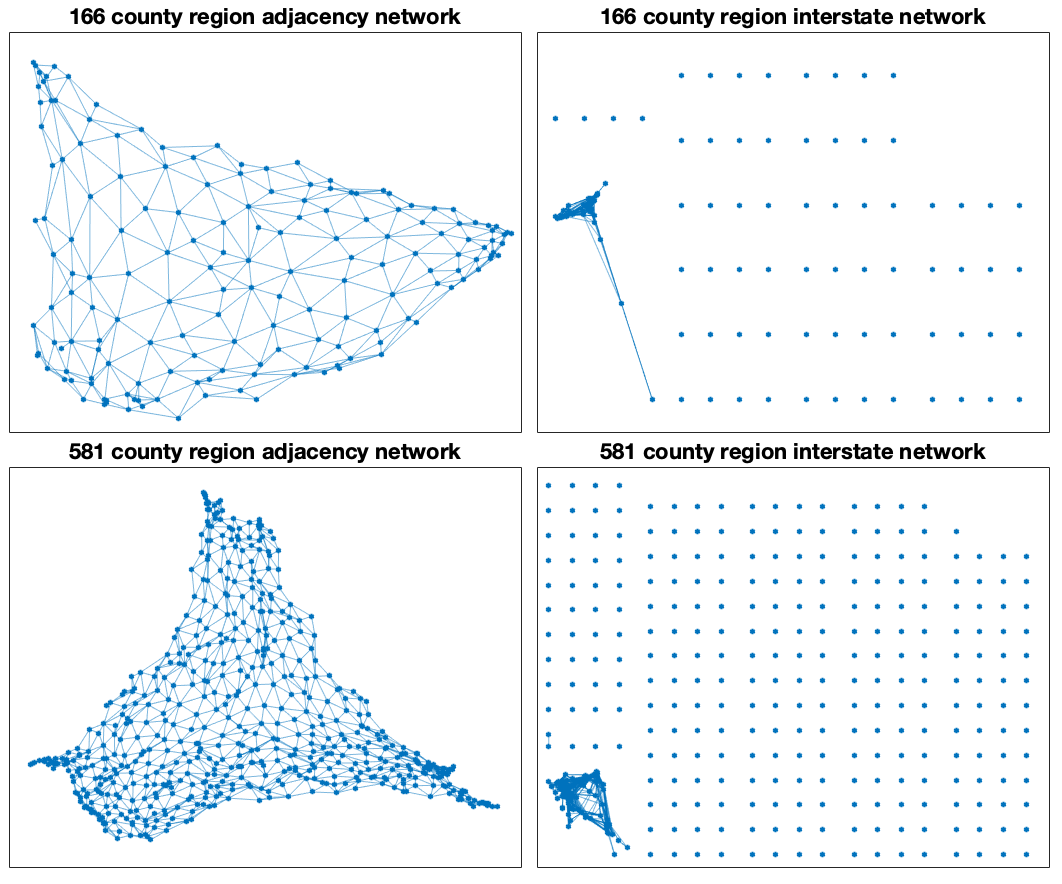}
\caption{ \textbf{The model networks}. The model space for the 166 county and the 581 county models consist of the combination of their adjacency and primary interstate networks. We note that the adjacency networks in both cases are fully connected, but that the primary interstate networks are not. \label{fig:networks}}

\end{figure}

The model space consists of two interconnected networks with the included counties as the nodes. See Fig \ref{fig:networks}. The adjacency matrices for the adjacent counties networks ($A$) were constructed using data from \cite{USCB2020}. The adjacency matrices for the primary interstate highway networks ($H$) were created using collected data on which primary interstates transect each county. 

\subsubsection*{Model parameterization}
We fit the parameter $p$ that represents the probability that an uninfested county adjacent to an infested county becomes infested from one year to the next year using the infestation data for 2014-2020 and following the approach in \cite{Leung2004}. More specifically, we proceed in the following steps. \\
1. Find the 'time $t$ to time $t+1$' infestation distribution as a function of infested adjacent counties using the 2014-2020 data. For all counties that are not infested at time $t$ count the number of infested adjacent counties $N_i$ they have and calculate the proportion $p_I$ of the counties that got infested in the next time step $t+1$.\\
2. Use the result from step 1 to fit the parameter $a$ in $p_I=1-e^{-a N_i}$. The Matlab function \texttt{fit} was used for this and the result was $a=0.2902$ with $95\%$ confidence interval $(0.2048,0.3756)$.\\
3. Use $a$ to calculate the parameter $p$ via $p=1-e^{-a}$ which yields $p=0.2519$.

\subsubsection*{Model implementation}
See Table \ref{tab:1} for the symbols and description of the model components. Each simulation starts with Berks county, PA, as the only infested county in 2014. Berks county is county number 76 in our 166 county data so at time 0 (2014) the infestation vector, which is a 166 element column vector, contains only 0's except for the 76th element which is a 1. Then on each timestep $t$ from 1 to $T$ the infestation vector $\bar{q}_t$ is updated by applying equations \ref{eq:1}-\ref{eq:3} and adding the result as a new column in a matrix $Q$ via the following Matlab code\\

\noindent \texttt{Q=[q]; \%Initiate the matrix for collecting infested over time. \\
for t=1:T\\
\indent q=(1-(1-p).\^{}(A*q))>rand(N,1)-q; \% adjacent spread\\
\indent q=(1/2)*(1-(1-sg).\^{}(H*q))>rand(N,1)-q; \% garden c interstate spread\\
\indent q=(1/2)*(1-(1-sp).\^{}(H*q))>rand(N,1)-q; \% population interstate spread\\
\indent Q=[Q,q]; \%add updated infestation vector to matrix\\
\noindent end}\\    

The output $Q$ is a matrix that contains the infestation status of each county (rows) in each year from 2014 to 2014$+T$ (columns). \texttt{sg} is the garden center establishment probability vector ($\bar{s}_g$ in Table \ref{tab:2}) and \texttt{sp} is the population establishment probability vector ($\bar{s}_p$ in Table \ref{tab:2}). The reason for including the \texttt{-q} in the right hand sides of the implementation of each of the model equations is to ensure that counties that have once become infested stay infested. This works because if $q_i=1$ then the corresponding left hand side is always larger than the right hand side and county $i$ will remain infested. This reflects reality, as no county that has become infested has subsequently become uninfested during the current US infestation. 

\subsubsection*{Model simulation and assessment}
To assess the accuracy of the model in the 166 county region we ran 1000 simulations starting with Berks county, PA, as the only infested county in 2014 and recorded the proportion of simulations that yielded each county infested in each year 2015-2021. Model assessment was carried out qualitatively by comparing infestation maps with model forecast maps through 2014-21 and quantitatively by defining a county infested if more than half of the simulations resulted in that specific county being infested in 2021. The qualitative comparison focused on comparing the overall annual range of the infestation and on the empirically observed corridors of spread west into OH, southwest along the VA-WV border, and south through VA. The quantitative assessment consisted of calculating the proportion of counties in the assessment region (Fig \ref{fig:1}A) that the model accurately predicted the infestation status of. For example, if the 2021 data says that a county is infested and more than 500 (of 1000) model simulations resulted in it being infested that is a match, as is if both model and data says not infested. Any disagreement between model and data on a particular county was recorded as a mismatch.

Once the model had been assessed in the 166 county region we extended the region to include 581 counties and collected data for each of the factors included in the original 166 county model. To forecast the future spread in the 581 county region we used the same parameterized model and approach as for the simulation in the 166 county region, but simulated for a longer period of time (2014-25). We again started with Berks County, PA, as the only infested county in 2014 and ran 1000 simulations and collected the proportion of simulations that each county got infested in each year up to and including 2025.



%
%
%

\newpage

\section*{Supporting information} 
\paragraph*{S1 Data.}
 \label{S1_Data}
{\bf Data and simulation results.} This Excel file contains all the data collected and generated in this study. It contains 4 tabs; 166 Region Data, 581 Region Data, 166 Region Model and 581 Region Model. The two Data tabs contain the collected data for each county in the respective regions. Specifically, county name, FIPS, number of garden centers, number of primary interstate highways (and their numbers) and population. The Model tabs contain the simulation outputs used to obtain the results presented in the manuscript with interactive maps. Readers can use these to run their own new simulations in Matlab (with or without changes to the code), export the results to the corresponding tab in this Excel file and see the results in map form. The 166 Region Model sheet also contains the calculations for the quantitative results, i.e. the number of model-data matches, false positives and false negatives.

\end{document}